\documentclass{article}
\usepackage[utf8]{inputenc}
\usepackage{graphicx}
\usepackage[margin=1in]{geometry}
\usepackage{amsmath,amssymb,amsthm,verbatim}
\usepackage[ruled,vlined]{algorithm2e}
\usepackage{algorithmic}
\usepackage{hyperref}
\usepackage{thm-restate}

\newtheorem{theorem}{Theorem}
\newtheorem{lemma}[theorem]{Lemma}
\newtheorem{proposition}[theorem]{Proposition}
\newtheorem{corollary}[theorem]{Corollary}

\SetKwFor{RepTimes}{repeat}{times}{end}

\title{The Randomized Competitive Ratio of Weighted $k$-server\\
is at Least Exponential}
\author{
Nikhil Ayyadevara\\
IIT Delhi\\
nikhil.ayyadevara@gmail.com
\and
Ashish Chiplunkar\\
IIT Delhi\\
ashishc@iitd.ac.in
}
\date{}

\begin{document}

\maketitle

\begin{abstract}
The weighted $k$-server problem is a natural generalization of the $k$-server problem in which the cost incurred in moving a server is the distance traveled times the weight of the server. Even after almost three decades since the seminal work of Fiat and Ricklin (1994), the competitive ratio of this problem remains poorly understood even on the simplest class of metric spaces -- the uniform metric spaces. In particular, in the case of randomized algorithms against the oblivious adversary, neither a better upper bound that the doubly exponential deterministic upper bound, nor a better lower bound than the logarithmic lower bound of unweighted $k$-server, is known. In this paper, we make significant progress towards understanding the randomized competitive ratio of weighted $k$-server on uniform metrics. We cut down the triply exponential gap between the upper and lower bound to a singly exponential gap by proving that the competitive ratio is at least exponential in $k$, substantially improving on the previously known lower bound of about $\ln k$.
\end{abstract}

\section{Introduction}

The $k$-server problem of Manasse, McGeoch, and Sleator~\cite{ManasseMS_STOC88} is one of the cleanest, simple-looking, and yet profound problems in online computation, and has been actively studied for over three decades. The $k$-server problem concerns deciding movements of $k$ mobile servers on an underlying metric space to serve a sequence of online requests. Each request is issued at some point of the metric space, and to serve such a request, a server must move to the requested point (unless a server is already present there). The cost incurred in the movement of a server is equal to the distance through which the server moves, and the goal is to minimize the total cost.

Since an online algorithm is required to take its decisions only based on the past inputs, it cannot output the optimal solution, in general. An online algorithm for a minimization problem is said to be $c$-\textit{competitive} if, on any instance, it produces a solution whose (expected) cost is at most $c$ times the cost of the optimum solution. The competitive ratio of an algorithm is the minimum (technically, the infimum of all) $c$ such that the algorithm is $c$-competitive. The deterministic (resp.\ randomized) competitive ratio of an online minimization problem is the minimum (technically, the infimum of all) $c$ for which a $c$-competitive deterministic (resp.\ randomized) algorithm exists. Note that, unless otherwise specified, we assume that in case of randomized algorithms, the adversarial input is \textit{oblivious}, that is, constructed with the knowledge of the algorithm but not the random choices the algorithm makes.

In their seminal work, Manasse, McGeoch, and Sleator~\cite{ManasseMS_STOC88} proved that the deterministic competitive ratio of the $k$-server problem is at least $k$ on every metric with more than $k$ points. They conjectured that the deterministic competitive ratio is, in fact, equal to $k$ on any metric. This conjecture is popularly called the deterministic $k$-server conjecture and it remains unresolved to date. The deterministic algorithm with the best known competitive ratio of $2k-1$ is due to Koutsoupias and Papadimitriou~\cite{KoutsoupiasP_JACM95}. Surprisingly, no better algorithm is known even using randomization. The randomized $k$-server conjecture states that a randomized algorithm with competitive ratio $O(\log k)$ exists on all metrics, and this remains unresolved after some recent progress~\cite{BubeckCLLM_STOC18,Lee_FOCS18}. The $k$-server problem on uniform metric spaces is particularly interesting because it is equivalent to the paging problem. In this case, several deterministic algorithms including Least-Recently-Used (LRU) and First-In-First-Out (FIFO) are known to be $k$-competitive. The randomized competitive ratio is known to be exactly $H(k)=\sum_{i=1}^k1/i\approx\ln k$, where the lower bound is due to Fiat et al.~\cite{FiatKLMSY_JAlg91} and the upper bound is due to \cite{McGeochS_Algorithmica91,AchlioptasCN_TCS00}.

The weighted $k$-server problem is a natural generalization of the $k$-server problem where the objective is to minimize the weighted sum of the movements of servers. Specifically, the $k$ servers have weights $\beta_1\leq\cdots\leq\beta_k$, and the cost of moving the $i$'th server is $\beta_i$ times the distance through which it moves. It is easy to see that a $c$-competitive $k$-server algorithm has competitive ratio at most $c\beta_k/\beta_1$ for the weighted $k$-server problem, and therefore, the challenge is to design an algorithm with competitive ratio independent of the servers' weights. Surprisingly, this innocuous-looking introduction of weights into the $k$-server problem makes it incredibly difficult, and a competitive algorithm is known only for $k\leq2$ \cite{Sitters_SIAMJC14} (of which, the $k=1$ case is trivial).

\subsection{Weighted $k$-Server on Uniform Metrics}

Owing to the difficulty of the weighted $k$-server problem on general metrics, the problem becomes particularly interesting on uniform metrics. The weighted $k$-server problem on uniform metric spaces models the paging problem where the cost of page replacement is determined by the location where the replacement takes place. Note that this problem is different from weighted caching~\cite{Young_Algorithmica02}, where the cost of page replacement is determined by the pages that get swapped in and out.

The seminal paper of Fiat and Ricklin~\cite{FiatR_TCS94} gave a deterministic algorithm for weighted $k$-server on uniform metrics whose competitive ratio is doubly exponential in $k$: about $3^{4^k/3}$ specifically, but can be improved to $2^{2^{k+2}}=16^{2^k}$ due to the result of Bansal et al.~\cite{BansalEKN_SODA18} for a more general problem. The fact that the deterministic competitive ratio is indeed doubly exponential in $k$ was established only recently by Bansal et al.~\cite{BansalEK_FOCS17}, who proved a lower bound of $2^{2^{k-4}}$, improving the previously known lower bound of $(k+1)!/2$ due to Fiat and Ricklin~\cite{FiatR_TCS94}.

The only known algorithm for the weighted $k$-server problem on uniform metrics which makes non-trivial use of randomness is by Chiplunkar and Vishwanathan~\cite{ChiplunkarV_TAlg20}. This algorithm also achieves a doubly exponential competitive ratio of about $c^{2^k}$ for $c\approx1.59792$. It is, in fact, a \textit{randomized memoryless algorithm} generalizing the algorithm by Chrobak and Sgall~\cite{ChrobakS_TCS04} for $k=2$, and it achieves the competitive ratio against a stronger form of adversary called \textit{adaptive online adversary}\footnote{An adaptive online adversary can see the movements of the algorithm's servers even though the algorithm is randomized. However, the adversary must also serve its requests in an online manner. The algorithm's cost is compared with the cost of the adversary's online solution to determine the competitive ratio.}. Chiplunkar and Vishwanathan also proved that no randomized memoryless algorithm can achieve a better competitive ratio against adaptive online adversaries. However, even when an algorithm is allowed to use both memory and randomness, and the adversary is oblivious, no better upper bound is known. More embarrassingly, for randomized algorithms, no better lower bound than the logarithmic lower bound of (unweighted) $k$-server on uniform metrics is known, thus, leaving a triply exponential gap between the upper and lower bounds.

In this paper, we cut down the triply exponential gap between the best known bounds on the randomized competitive ratio of weighted $k$-server on uniform metrics by a doubly exponential improvement in the lower bound. We prove,

\begin{theorem}\label{thm_main_abstract}
The competitive ratio of any randomized algorithm for weighted $k$-server on uniform metrics is at least exponential in $k$, even when the algorithm is allowed to use memory and the adversary is oblivious.
\end{theorem}

Due to our result, we now have only a singly exponential gap between the best known upper and lower bounds on the randomized competitive ratio of weighted $k$-server on uniform metrics.

\subsection{Comparison with the Deterministic Lower Bound}

Our proof of the randomized lower bound for weighted $k$-server is inspired by the proof of the deterministic lower bound by Bansal et al.~\cite{BansalEK_FOCS17}. Both proofs give adversaries which run recursively defined strategies relying crucially on a certain set-system $\mathcal{Q}$. However, our proof differs in the following aspects.
\begin{enumerate}
\item The adversary in the deterministic lower bound proof is able to carefully pick from $\mathcal{Q}$ a set of points that does not contain points covered by the algorithm's heavier servers, and run its strategy on that set. In contrast, our adversary is oblivious and is unable to see the positions of the algorithm's servers. Therefore, it merely picks a random set from $\mathcal{Q}$ and hopes that none of the points in that set is covered by the algorithm's heavier servers.
\item The strategy of Bansal et al.\ to defeat deterministic algorithms ensures that whenever an adversary's server other than the heaviest moves, it is accompanied by an eventual movement of a heavier server of the algorithm. Therefore, assuming that the weights of the servers are well-separated, their task reduces to proving that the heaviest server of the algorithm moves a large number of times as compared to the heaviest server of the adversary. On the other hand, we are unable to charge the movement of an adversary's server to the movement of an algorithm's heavier server. Consequently, we need to carefully track the contributions of all $k$ servers towards the algorithm's and the adversary's costs.
\end{enumerate}

\section{Preliminaries}

Let the weights of the $k$ servers be $1,\beta,\beta^2,\ldots,\beta^{k-1}$ for some large integer $\beta$ which we will fix later. Define the sequence $n_0,n_1,\ldots$ inductively as follows. $n_0=1$, and for $\ell>0$,
\[n_{\ell}=\left(\left\lceil\frac{n_{\ell-1}}{2}\right\rceil+1\right)\cdot\left(\left\lfloor\frac{n_{\ell-1}}{2}\right\rfloor+1\right)\text{.}\]
Observe that $n_k$ grows doubly exponentially with respect to $k$. Since $n_{\ell}\geq n_{\ell-1}^2/4$, it is easy to prove using induction that $n_{\ell}\geq4\cdot(64^{1/32})^{2^{\ell}}$ for all $\ell\geq5$. Let $H$ denote the harmonic function, that is, $H(n)=\sum_{i=1}^n1/i$. It is known that $H(n)\geq\ln n$. We will establish Theorem~\ref{thm_main_abstract} by proving the following bound.

\begin{restatable}{theorem}{mainconcrete}\label{thm_main_concrete}
The randomized competitive ratio of weighted $k$-server on uniform metric spaces is at least $H(n_{k-1})=\Omega(2^k)$.
\end{restatable}

We use the following version of Yao's principle to prove the above bound.

\begin{proposition}[Yao's principle]\label{prop_Yao}
Suppose there exists a probability distribution $\mathcal{D}$ on the instances of an online minimization problem such that for every deterministic online algorithm $A$, we have,
\[\mathbb{E}_{I\sim\mathcal{D}}[A(I)]>\alpha\cdot\mathbb{E}_{I\sim\mathcal{D}}[\text{OPT}(I)]\text{,}\]
where $A(I)$ is the cost of the algorithm's solution and $\text{OPT}(I)$ is the cost of an optimal solution to instance $I$. Then the problem does not have an $\alpha$-competitive randomized online algorithm.
\end{proposition}

Thus, in order to prove Theorem~\ref{thm_main_concrete}, our task is exhibit a distribution on instances of weighted $k$-server on a uniform metric space such that the expected cost of any deterministic online algorithm is greater than $H(n_{k-1})$ times the expectation of the optimum cost. To construct our distribution on instances, we use a combinatorial result with a constructive proof given by Bansal et al.~\cite{BansalEK_FOCS17}. We reproduce its proof in Appendix~\ref{app_setsystem} for completeness. The result is as follows.

\begin{restatable}{lemma}{setsystem}\label{lem_setsystem}
Let $\ell\in\mathbb{N}$ and let $P$ be a set of $n_{\ell}$ points. There exists a set-system $\mathcal{Q}_{\ell}\subseteq2^P$ satisfying the following properties.
\begin{enumerate}
\item $\mathcal{Q}_{\ell}$ contains $\lceil n_{\ell-1}/2\rceil+1$ sets, each of size $n_{\ell-1}$.
\item For every $p\in P$, there exists a set in $\mathcal{Q}_{\ell}$ not containing $p$.
\item For every $p\in P$, there exists a $q\in P$ such that every set in $\mathcal{Q}_{\ell}$ contains at least one of $p$ and $q$.
\end{enumerate}
\end{restatable}

\section{Adversarial Strategy and Analysis}

Consider the uniform metric space on a set $S$ of $n_{k-1}+1$ points. Our adversarial input distribution is generated by the procedure \textsf{adversary} which uses a recursive procedure \textsf{strategy}, an oblivious version of its counterpart in Bansal et al.~\cite{BansalEK_FOCS17}. These procedures are defined as follows.

\begin{algorithm}[H]
\SetAlgorithmName{Procedure}
\renewcommand{\thealgorithm}{}
\caption{\textsf{adversary}}
\RepTimes{infinitely many}{
Pick a point $p$ uniformly at random from $S$ (with replacement)\;
Call \textsf{strategy}$(k-1,S\setminus\{p\})$\;
}
\end{algorithm}

\begin{algorithm}[H]
\SetAlgorithmName{Procedure}
\renewcommand{\thealgorithm}{}
\caption{\textsf{strategy}$(\ell,P)$ (Promise: $|P|=n_{\ell}$)}
\eIf{$\ell=0$ (and therefore, $|P|=n_0=1$)}{
Request the unique point in $P$\;
}{
Construct the set-system $\mathcal{Q}_{\ell}\subseteq2^P$ using Lemma~\ref{lem_setsystem}\;
\RepTimes{$(\beta-1)\cdot\left(\lceil n_{\ell-1}/2\rceil+1\right)$}{
Pick a set $P'$ uniformly at random from $\mathcal{Q}_{\ell}$ (with replacement)\;
Call \textsf{strategy}$(\ell-1,P')$\;
}
}
\end{algorithm}

Procedure \textsf{strategy} gets as input a non-negative number $\ell$ and a set $P$ of $n_{\ell}$ points. In the base case where $\ell=0$, the procedure issues a request to the unique point in $P$. In the inductive case where $\ell>0$, the procedure constructs the set-system $\mathcal{Q}_{\ell}$ with properties stated in Lemma~\ref{lem_setsystem} on the set $P$. Then it repeatedly gives recursive calls, passing $\ell-1$ in place of $\ell$, on sets chosen uniformly at random from $\mathcal{Q}_{\ell}$. Recall that these sets have size $n_{\ell-1}$, as required. Procedure \textsf{adversary} takes a uniform metric space on $n_{k-1}+1$ points. It repeatedly picks a point $p$ uniformly at random and calls the procedure \textsf{strategy} on the set of points other than $p$.

For analysis, fix an arbitrary deterministic online algorithm and the initial positions of its servers. We first consider requests given by one execution of procedure \textsf{strategy}$(\ell,P)$, and bound the number of movements of the algorithm's servers to serve those requests.

\begin{restatable}{lemma}{algrec}\label{lem_alg_rec}
For every $\ell\in\{0,\ldots,k-1\}$ the following holds. Let $\rho_0$ be an arbitrary sequence of requests and $L$ be the set of positions of the algorithm's heaviest $k-\ell$ servers after serving $\rho_0$. Let $P$ be an arbitrary set of $n_{\ell}$ points disjoint from $L$. Suppose $\rho_0$ is followed by a random sequence $\rho$ of requests given by a \textsf{strategy}$(\ell,P)$ call. For $i=1,\ldots,k$, let the random variable $X_i$ denote the number of movements of the algorithm's $i$'th lightest server while the algorithm serves $\rho$. Then we have,
\[\sum_{i=1}^k\beta^{\min(i-1,\ell)}\cdot\mathbb{E}[X_i]\geq(\beta-1)^{\ell}\text{.}\]
\end{restatable}

We defer the proof of this lemma to Appendix~\ref{app_alg}. On a high level, the proof goes as follows. If the algorithm moves one of its heaviest $k-\ell$ servers while it serves $\rho$, then it pays a lot already. If not, it must serve $\rho$ using its lightest $\ell$ servers only. In this case, each recursive call given by the \textsf{strategy}$(\ell,P)$ call is, with sufficient probability, on a set $P'$ not containing the location of the algorithm's $\ell$'th lightest server. This enables us to use induction hypothesis to bound the algorithm's cost in each recursive call.

Intuitively, Lemma~\ref{lem_alg_rec} gives a lower bound of $(\beta-1)^{\ell}$ on the expected cost incurred by the algorithm in serving requests given by a \textsf{strategy}$(\ell,P)$ call, but with the following caveat: movements of the heaviest $k-\ell-1$ servers are charged at a discounted rate of $\beta^{\ell}$. However, when $\ell$ is instantiated to $k-1$ in particular, no discount remains applicable. Thus, $(\beta-1)^{k-1}$ becomes a lower bound on the actual expected cost of the algorithm in serving requests given by a \textsf{strategy}$(k-1,P)$ call. With this observation, we immediately get the following bound on the expected cost of the algorithm in serving requests given by each \textsf{strategy} call made by the procedure \textsf{adversary}.

\begin{corollary}[to Lemma~\ref{lem_alg_rec}]\label{cor_alg}
The expected cost of the algorithm in serving requests given by each \textsf{strategy} call made by  \textsf{adversary} is at least $(\beta-1)^{k-1}/(n_{k-1}+1)$.
\end{corollary}

\begin{proof}
Consider any \textsf{strategy}$(k-1,S\setminus\{p\})$ call, where $p$ is a uniformly random point in $S$. Let $r$ be the location of the algorithm's heaviest server at the time the call is made. Then $\Pr[r\notin S\setminus\{p\}]=\Pr[p=r]=1/|S|=1/(n_{k-1}+1)$. Lemma~\ref{lem_alg_rec} implies that conditioned on $r\notin S\setminus\{p\}$, the expected cost of the algorithm in serving requests given by the \textsf{strategy}$(k-1,S\setminus\{p\})$ call is at least $(\beta-1)^{k-1}$. Thus, the claim follows.
\end{proof}

Let us now turn our attention towards the adversary's cost. We will show how the adversary, having the ability to see the future requests, can ensure that whenever \textsf{strategy}$(\ell,P)$ is called, it has at least one server other than its $\ell$ lightest servers occupying a point in $P$ already. On the contrary, recall that in Lemma~\ref{lem_alg_rec}, we relied on the algorithm not having any of its servers except the $\ell$ lightest ones occupying points in $P$ at the time \textsf{strategy}$(\ell,P)$ is called. Intuitively, the adversary is able to obtain advantage over the algorithm by having one server other than the $\ell$ lightest ones in $P$ whereas the algorithm has none.

\begin{lemma}\label{lem_adv_rec}
Define the sequence $c_0,c_1,\ldots$ inductively as follows: $c_0=0$, and for $\ell>0$,
\[c_{\ell}=\beta^{\ell-1}+\beta\cdot\left(\lceil n_{\ell-1}/2\rceil+1\right)\cdot c_{\ell-1}\text{.}\]
Suppose that the adversary has at least one server other than its $\ell$ lightest servers occupying some point in $P$ at the time \textsf{strategy}$(\ell,P)$ is called. Then the adversary is able to serve all requests given in this call with cost at most $c_{\ell}$ by moving only its $\ell$ lightest servers.
\end{lemma}

\begin{proof}
We prove the claim by induction on $\ell$. For the base case, suppose $\ell=0$. Then $|P|=1$ and by assumption, the adversary has at least one server at the unique point in $P$. Therefore, the adversary can serve the unique request given by \textsf{strategy}$(0,P)$ with cost $c_0=0$, without moving any server.

For the inductive case, suppose $\ell>0$. We have assumed that the adversary has at least one server other than its lightest $\ell$ servers occupying some point $p$ in $P$. By the third property of the set-system $\mathcal{Q}_{\ell}$ from Lemma~\ref{lem_setsystem}, there exists a point $q\in P$ such that each set in $\mathcal{Q}_{\ell}$ contains at least one of $p$ and $q$. The adversary moves its $\ell$'th lightest server to such a point $q$ and keeps it there until the end of the \textsf{strategy}$(\ell,P)$ call. Due to this movement, the adversary incurs cost $\beta^{\ell-1}$, the first term in the definition of $c_{\ell}$. As a result, both $p$ and $q$ become occupied by the adversary's servers other than the $\ell-1$ lightest ones. We now show how the requests in all recursive calls made by \textsf{strategy}$(\ell,P)$ can be served by moving the $\ell-1$ lightest servers only.

Consider any of the recursive calls made by \textsf{strategy}$(\ell,P)$. The set $P'\in\mathcal{Q}_{\ell}$ on which this call is made contains at least one of $p$ and $q$. Both $p$ and $q$ were occupied by the adversary's servers other than the $\ell-1$ lightest ones before \textsf{strategy}$(\ell,P)$ made its first recursive call. All the previous recursive calls were served by moving only the $\ell-1$ lightest servers. Thus, at the time the current recursive call \textsf{strategy}$(\ell-1,P')$ is made, points $p$ and $q$ are still occupied  by the adversary's servers other than the $\ell-1$ lightest ones. Therefore, at least one of these servers occupies a point in $P'$. By induction hypothesis, the adversary can serve all requests in the current recursive call \textsf{strategy}$(\ell-1,P')$ with cost at most $c_{\ell-1}$ by moving only the $\ell-1$ lightest servers. Since the number of such recursive calls is $(\beta-1)\cdot\left(\lceil n_{\ell-1}/2\rceil+1\right)\leq\beta\cdot\left(\lceil n_{\ell-1}/2\rceil+1\right)$, the adversary serves all requests made in these calls with cost at most $\beta\cdot\left(\lceil n_{\ell-1}/2\rceil+1\right)\cdot c_{\ell-1}$, the second term in the expression for $c_{\ell}$.
\end{proof}

We now use Corollary~\ref{cor_alg} and Lemma~\ref{lem_adv_rec} to prove Theorem~\ref{thm_main_concrete}.

\mainconcrete*

\begin{proof}
We track the costs incurred by the algorithm and the adversary per \textsf{strategy}$(k-1,P)$ call made by the procedure \textsf{adversary}, and show that the former is at least $H(n_{k-1})$ times the latter.

Here is how the adversary serves the requests. Let $q$ denote the position of the adversary's heaviest server at the time a \textsf{strategy}$(k-1,P)$ call is made. If $P=S\setminus\{q\}$, that is, the random point sampled from $S$ turns out to be $q$, then the adversary finds the point $q'$ which is sampled farthest in future by the procedure \textsf{adversary}, and moves its heaviest server there. These are the only movements of the adversary's heaviest server. By the standard coupon-collector argument, the expected number of samples from the current sample of $q$ to $q'$ is $(n_{k-1}+1)H(n_{k-1})$, because $|S|=n_{k-1}+1$ and we have already sampled $q$. Thus, in the long run, the cost of the adversary resulting from moving its heaviest server, per \textsf{strategy} call made by the procedure \textsf{adversary}, is $\beta^{k-1}/((n_{k-1}+1)H(n_{k-1}))$.

By moving its heaviest server as described above, the adversary ensures the following. Before the adversary starts serving requests given by a \textsf{strategy}$(k-1,S\setminus\{p\})$ call, its heaviest server is located at some point different from $p$, and therefore, in $S\setminus\{p\}$. By Lemma~\ref{lem_adv_rec}, the adversary is able to serve requests given by each \textsf{strategy}$(k-1,S\setminus\{p\})$ with cost at most $c_{k-1}$ without moving its heaviest server. In other words, the contribution of the adversary's servers other than the heaviest towards its cost per \textsf{strategy} call is at most $c_{k-1}$.

Thus, the adversary's cost per \textsf{strategy} call made by the procedure \textsf{adversary} is at most $\beta^{k-1}/((n_{k-1}+1)\cdot H(n_{k-1}))+c_{k-1}$, which, by unrolling the recurrence in the statement of Lemma~\ref{lem_adv_rec}, is given by
\[\frac{\beta^{k-1}}{(n_{k-1}+1)\cdot H(n_{k-1})}+c_{k-1}=\frac{\beta^{k-1}}{(n_{k-1}+1)\cdot H(n_{k-1})}+\beta^{k-2}\cdot\sum_{i=1}^{k-1}\prod_{j=i}^{k-2}\left(\left\lceil\frac{n_j}{2}\right\rceil+1\right)\text{.}\]
Let $\varepsilon$ be an arbitrarily small positive number. By choosing
\[\beta=\lceil\varepsilon^{-1}\rceil\cdot(n_{k-1}+1)\cdot H(n_{k-1})\cdot\sum_{i=1}^{k-1}\prod_{j=i}^{k-2}\left(\left\lceil\frac{n_j}{2}\right\rceil+1\right)\text{,}\]
the adversary's cost per \textsf{strategy} call is bounded from above by
\[\frac{\beta^{k-1}\cdot(1+\varepsilon)}{(n_{k-1}+1)\cdot H(n_{k-1})}\text{.}\] 

On the other hand, recall from Corollary~\ref{cor_alg} that the expected cost of the algorithm per \textsf{strategy} call made by the procedure \textsf{adversary} is at least
\[\frac{(\beta-1)^{k-1}}{n_{k-1}+1}\geq\frac{\beta^{k-1}}{n_{k-1}+1}\cdot\left(1-\frac{1}{\beta}\right)^{k-1}\geq\frac{\beta^{k-1}}{n_{k-1}+1}\cdot\left(1-\frac{k-1}{\beta}\right)\geq\frac{\beta^{k-1}\cdot(1-\varepsilon)}{n_{k-1}+1}\text{,}\]
because $\beta\gg k/\varepsilon$. Thus, modulo the $(1\pm\varepsilon)$ factors, the algorithm's cost per \textsf{strategy} call is at least $H(n_{k-1})$ times the adversary's cost per \textsf{strategy} call. Since $\varepsilon$ is arbitrarily small, we use Proposition~\ref{prop_Yao} to conclude that the competitive ratio of any randomized online algorithm for weighted $k$-server on uniform metrics is at least $H(n_{k-1})$.
\end{proof}

\section{Concluding Remarks}

Given our lower bound on the randomized competitive ratio of weighted $k$-server on uniform metric spaces, the gap between the known upper and lower bounds has reduced from three orders of exponentiation to one. The natural question that needs to be investigated is to determine the randomized competitive ratio, or at least, prove upper and lower bounds that match in the order of exponentiation.

Our result also sheds light on the randomized competitive ratio of a generalization of the weighted $k$-server problem on uniform metrics called the generalized $k$-server problem on weighted uniform metrics. In this problem $k$ servers are restricted to move in $k$ different uniform metric spaces that are scaled copies of one another. A request contains one point from each copy and to serve it, one of the points must be covered by the server moving in its copy. Our lower bound directly applies to the generalized $k$-server problem on weighted uniform metrics and improves the previously known lower bound\footnote{This bound, in fact, holds for the unweighted counterpart, and to the best of the authors' knowledge, no better bound for the weighted problem was known.} of $\Omega(k/\log^2k)$ by Bansal et al.~\cite{BansalEKN_SODA18} to exponential in $k$. This also proves that the generalized $k$-server problem on weighted uniform metrics is qualitatively harder than its unweighted counterpart, the generalized $k$-server problem on uniform metrics, which has randomized competitive ratio $O(k^2\log k)$ due to Beinkowski, Je\.{z}, and Schmidt~\cite{BienkowskiJS_ISAAC19}.

\bibliographystyle{plain}
\bibliography{references.bib}

\appendix

\section{Set-system Construction}\label{app_setsystem}

\setsystem*

\begin{proof}[Proof (Bansal et al.~\cite{BansalEK_FOCS17})]
Construct the set-system $\mathcal{Q}_{\ell}$ as follows. Recall that
\[|P|=n_{\ell}=\left(\left\lceil\frac{n_{\ell-1}}{2}\right\rceil+1\right)\cdot\left(\left\lfloor\frac{n_{\ell-1}}{2}\right\rfloor+1\right)\text{.}\]
Let $M$ be an arbitrary subset of $P$ having size $\left\lceil n_{\ell-1}/2\right\rceil+1$, so that
\[|P\setminus M|=\left(\left\lceil\frac{n_{\ell-1}}{2}\right\rceil+1\right)\cdot\left\lfloor\frac{n_{\ell-1}}{2}\right\rfloor\text{.}\]
Partition $P\setminus M$ into $\left\lceil n_{\ell-1}/2\right\rceil+1$ sets of size $\left\lfloor n_{\ell-1}/2\right\rfloor$ each, and for each $r\in M$, name a distinct set in the partition $P'_r$. Next, for each $r\in M$, define $P_r=(M\setminus\{r\})\cup P'_r$, and let $\mathcal{Q}_{\ell}=\{P_r\text{ }|\text{ }r\in M\}$.

We now prove that $\mathcal{Q}_{\ell}$ indeed satisfies the required properties. First, the number of sets in $\mathcal{Q}_{\ell}$ is equal to $|M|=\left\lceil n_{\ell-1}/2\right\rceil+1$, and the size of each set $P_r\in\mathcal{Q}_{\ell}$ is
\[|P_r|=|M|-1+|P'_r|=\left\lceil\frac{n_{\ell-1}}{2}\right\rceil+\left\lfloor\frac{n_{\ell-1}}{2}\right\rfloor=n_{\ell-1}\text{.}\]
For the second property, observe that a point $p\in M$ is not contained in the corresponding set $P_p\in\mathcal{Q}_{\ell}$, whereas for a point $p\in P'_r$, the only set in $\mathcal{Q}_{\ell}$ that contains $p$ is $P_r$. For the third property, if $p\in M$, define $q$ to be any other point in $M$, and if $p\in P'_r$, define $q=r$, and check that the property is indeed satisfied. 
\end{proof}

\section{Analysis of the Algorithm's Movements}\label{app_alg}

We present the proof of Lemma~\ref{lem_alg_rec} here, for which we need the following lemma.

\begin{lemma}\label{lem_rv}
Let $Z_1$ and $Z_2$ be non-negative random variables and $E$ be an event on a common sample space such that $\mathbb{E}[Z_1\mid E]\geq b$ and $\mathbb{E}[Z_2\mid\neg E]\geq b$ for some real number $b$. Then $\mathbb{E}[Z_1+Z_2]\geq b$.
\end{lemma}

\begin{proof}
We have,
\[\mathbb{E}[Z_1+Z_2]=\mathbb{E}[Z_1+Z_2\mid E]\cdot\Pr[E]+\mathbb{E}[Z_1+Z_2\mid\neg E]\cdot\Pr[\neg E]\text{.}\]
Since $Z_1$ and $Z_2$ are non-negative, we have,
\[\mathbb{E}[Z_1+Z_2]\geq\mathbb{E}[Z_1\mid E]\cdot\Pr[E]+\mathbb{E}[Z_2\mid\neg E]\cdot\Pr[\neg E]\geq b\cdot\Pr[E]+b\cdot\Pr[\neg E]=b\text{,}\]
as required.
\end{proof}

\algrec*

\begin{proof}
We prove the claim by induction on $\ell$. For the base case, suppose $\ell=0$. Then $|P|=1$, and we are assured that $L$, the set of points occupied by all the algorithm's servers, is disjoint from $P$. In other words, none of the algorithm's servers occupy the unique point in $P$. Therefore, to serve the one request given by \textsf{strategy}$(0,P)$, the algorithm must move at least one of its servers, and thus,
\[\sum_{i=1}^k\beta^{\min(i-1,\ell)}\cdot\mathbb{E}[X_i]=\sum_{i=1}^k\mathbb{E}[X_i]\geq1=(\beta-1)^{\ell}\text{,}\]
as required.

For the inductive case, suppose $\ell>0$. We are assured that except for the lightest $\ell$ servers, none of the servers of the algorithm occupy points in $P$ at the time the \textsf{strategy}$(\ell,P)$ call is made. This call makes $m=(\beta-1)\cdot\left(\lceil n_{\ell-1}/2\rceil+1\right)=(\beta-1)\cdot|\mathcal{Q}_{\ell}|$ recursive calls. For $i=1,\ldots,k$ and $j=1,\ldots,m$, let the random variable $Y_i^j$ denote the number of movements of the algorithm's $i$'th lightest server to serve requests from the $j$'th recursive call. Thus, for all $i$, $X_i=\sum_{j=1}^mY_i^j$.

Consider an arbitrary $j\in\{1,\ldots,m\}$. Let $E_j$ denote the event that the random variables $Y_i^{j'}$ are all $0$ for all $i>\ell$ and $j'<j$. In words, $E_j$ is the event that none of the algorithm's heaviest $k-\ell$ servers move during the first $j-1$ recursive calls. Recall that originally these servers did not occupy any point in $P$. Therefore, if $E_j$ happens, these servers are guaranteed to be out of the set $P'\subseteq P$ on which the $j$'th recursive call is made. Next, let $E'_j$ denote the event that the $j$'th recursive call is made on a set $P'$ that does not contain the position of the algorithm's $\ell$'th lightest server after the first $j-1$ recursive calls. Thus, if both $E_j$ and $E'_j$ happen, then $P'$ is disjoint from the set of positions of the algorithm's $k-\ell+1$ heaviest servers. We can then apply the induction hypothesis to get,
\begin{equation}\label{eqn_alg_ind}
\sum_{i=1}^k\beta^{\min(i-1,\ell-1)}\cdot\mathbb{E}[Y_i^j\mid E_j\wedge E'_j]\geq(\beta-1)^{\ell-1}\text{.}
\end{equation}
Next, let us understand the behavior of the random variables $Y_i^j$ conditioned on $E_j$ only. We have,
\begin{equation}\label{eqn_inter1}
\mathbb{E}[Y_i^j\mid E_j]\geq\mathbb{E}[Y_i^j\mid E_j\wedge E'_j]\cdot\Pr[E'_j\mid E_j]\geq\frac{\mathbb{E}[Y_i^j\mid E_j\wedge E'_j]}{|\mathcal{Q}_{\ell}|}\text{.}
\end{equation}
Here, the first inequality holds because $Y_i^j$ is non-negative. The second inequality holds because, by the second property of the set-system $\mathcal{Q}_{\ell}$ given by Lemma~\ref{lem_setsystem}, for every possible history before the $j$'th recursive call, $\mathcal{Q}_{\ell}$ contains at least one set which does not contain the position of the algorithm's $\ell$'th lightest server. This implies $\Pr[E'_j\mid E_j]\geq1/|\mathcal{Q}_{\ell}|$. From Equation~\ref{eqn_alg_ind} and Equation~\ref{eqn_inter1}, we get,
\begin{equation}\label{eqn_inter2}
\sum_{i=1}^k\beta^{\min(i-1,\ell-1)}\cdot\mathbb{E}[Y_i^j\mid E_j]\geq\frac{(\beta-1)^{\ell-1}}{|\mathcal{Q}_{\ell}|}\text{.}
\end{equation}
By the non-negativity of the random variables $Y_i^{j'}$ and the definition of $E_j$, we trivially have,
\[\sum_{i=\ell+1}^k\sum_{j'=1}^m\mathbb{E}[Y_i^{j'}\mid\neg E_j]\geq\sum_{i=\ell+1}^k\sum_{j'=1}^{j-1}\mathbb{E}[Y_i^{j'}\mid\neg E_j]\geq1\text{,}\]
and hence,
\begin{equation}\label{eqn_trivial}
\frac{(\beta-1)^{\ell-1}}{|\mathcal{Q}_{\ell}|}\cdot\sum_{i=\ell+1}^k\sum_{j'=1}^m\mathbb{E}[Y_i^{j'}\mid\neg E_j]\geq\frac{(\beta-1)^{\ell-1}}{|\mathcal{Q}_{\ell}|}\text{.}
\end{equation}
From Equation~\ref{eqn_inter2} and Equation~\ref{eqn_trivial}, using Lemma~\ref{lem_rv}, we get,
\[\sum_{i=1}^k\beta^{\min(i-1,\ell-1)}\cdot\mathbb{E}[Y_i^j]+\frac{(\beta-1)^{\ell-1}}{|\mathcal{Q}_{\ell}|}\cdot\sum_{i=\ell+1}^k\sum_{j'=1}^m\mathbb{E}[Y_i^{j'}]\geq\frac{(\beta-1)^{\ell-1}}{|\mathcal{Q}_{\ell}|}\text{.}\]

The above inequality holds for all $j\in\{1,\ldots,m\}$. Summing up over all $j$ and recalling $X_i=\sum_{j=1}^mY_i^j$, we get,
\[\sum_{i=1}^k\beta^{\min(i-1,\ell-1)}\cdot\mathbb{E}[X_i]+m\cdot\frac{(\beta-1)^{\ell-1}}{|\mathcal{Q}_{\ell}|}\cdot\sum_{i=\ell+1}^k\mathbb{E}[X_i]\geq m\cdot\frac{(\beta-1)^{\ell-1}}{|\mathcal{Q}_{\ell}|}\text{.}\]
Recall that $m=(\beta-1)\cdot|\mathcal{Q}_{\ell}|$. Thus,
\begin{equation}\label{eqn_inter3}
\sum_{i=1}^k\beta^{\min(i-1,\ell-1)}\cdot\mathbb{E}[X_i]+(\beta-1)^{\ell}\cdot\sum_{i=\ell+1}^k\mathbb{E}[X_i]\geq(\beta-1)^{\ell}\text{.}
\end{equation}
Finally, note that for $i>\ell$, $\min(i-1,\ell-1)=\ell-1$, and since $\ell\geq1$, we have $(\beta-1)^{\ell}\leq\beta^{\ell-1}(\beta-1)$. Therefore, the multiplier of the $\mathbb{E}[X_i]$ term in Equation~\ref{eqn_inter3} is bounded as,
\[\beta^{\min(i-1,\ell-1)}+(\beta-1)^{\ell}\leq\beta^{\ell-1}+\beta^{\ell-1}(\beta-1)=\beta^{\ell}=\beta^{\min(i-1,\ell)}\text{.}\]
On the other hand, for $i\leq\ell$, $\min(i-1,\ell-1)=\min(i-1,\ell)$. Thus, for all $i\in\{1,\ldots,k\}$, the multiplier of the $\mathbb{E}[X_i]$ term in Equation~\ref{eqn_inter3} is at most $\beta^{\min(i-1,\ell)}$. Since the $X_i$'s are all non-negative, we get,
\[\sum_{i=1}^k\beta^{\min(i-1,\ell)}\cdot\mathbb{E}[X_i]\geq(\beta-1)^{\ell}\text{,}\]
as required.
\end{proof}

\end{document}